\documentclass{paper}
\usepackage{apacite}
\usepackage{graphicx}
\usepackage{dirtytalk}
\usepackage{etoolbox}
\usepackage{placeins}
\usepackage[rightcaption]{sidecap}
\usepackage[]{algorithm2e}

\AtBeginEnvironment{quote}{\small}

\begin{document}

\title{On the marginal utility of fiat money: insurmountable circularity or not?}

\author{Michael Reiss \\ \small{reissgo@gmail.com}}
\maketitle

\begin{abstract}
The question of how a pure fiat currency is enforced and comes to have a non-zero value has been much debated \cite{selgin1994}. What is less often addressed is the case where the enforcement is taken for granted and we ask what value (in terms of goods and services) the currency will end up taking. Establishing a decentralised mechanism for price formation has proven a challenge for economists:
\begin{quote}
Since no decentralized out-of-equilibrium adjustment mechanism has been discovered, we currently have no acceptable dynamical model of the Walrasian system.\cite{gintis2006emergence}

\end{quote}
In his paper, Gintis put forward a model for price discovery based on the evolution of the model's agents, i.e. ``poorly performing agents dying and being replaced by copies of the well-performing agents.'' It seems improbable that this mechanism is the driving force behind price discovery in the real world. This paper proposes a more realistic mechanism and presents results from a corresponding agent-based model.

\end{abstract}
\section{Introduction}

The utility of some good is an expression of how much wellbeing it would bring us. A typical utility function for some good X may look like figure 1 where you would expect a diminishing returns effect, whereby the nth unit of good X brings less additional utility than the (n-1)th  unit of good X.

\begin{figure}

\includegraphics[width=9cm]{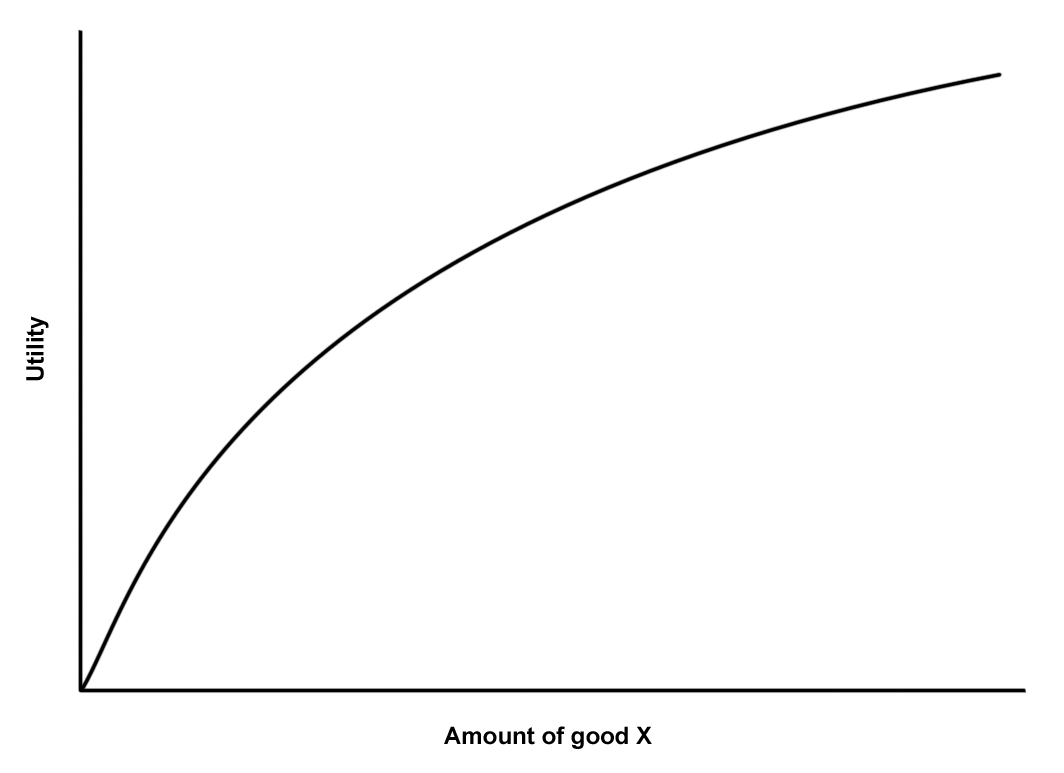}
\caption{Typical utility function for a good}
\end{figure}

Let us now consider what such a utility curve may look like for money itself – specifically fiat money with no intrinsic value. One might imagine that estimating such a curve, at least in theory, would be relatively straightforward. People already know the price of a wide variety of goods as their prices are available for all to see, so people know what their money can buy. In this case the marginal utility of money becomes simply the marginal utility of whatever goods an agent wishes to purchase with that money. But now consider what the situation would be if people did not know the typical prices of goods, what if we had a theoretical, pre-money, barter economy and a government abruptly issued everyone with some collection of say, “shillings” and instructed the population to start buying and selling goods with the new currency. Under this scenario the population is in a seemingly impossible situation. How many shillings you would be willing to sell your produce for would depend on how much utility you would gain by receiving the payment but we cannot estimate that until we know how much we could buy with the shillings. This circularity problem with fiat money has been recognised by many economists, for example:

\begin{quote}
It was not long until the entire theory of marginal utility was abandoned, since it obviously rested on circular reasoning. Although it tried to explain prices, prices were necessary to explain marginal utility. \cite{70488}
\end{quote}

and

\begin{quote}To determine the price of a good, we analyze the market ­demand schedule for the good; this, in turn, depends on the individual demand schedules; these in their turn are determined by the individuals’ value rankings of units of the good and units of money as given by the various alternative uses of money; yet the latter alternatives depend in turn on given prices of the other goods,
\cite{rothbard2009man}\end{quote}

One may observe that whatever the issue might be in a theoretical barter transition to fiat currency scenario, the problem does not appear to exist in the real world today. The price of items has somehow become established and, in the short term at least, appears reasonably stable. There are however, still good reasons to obtain a theoretical understanding of the marginal utility of money in order to better understand the effects of the many continuous changes in the economy. For example, changes in production efficiency, supply shocks or indeed changes in the supply of money. Any time there is a change in any of these factors, it will undoubtedly change people’s marginal utility of money. If there were a series of technological advances that made the production of goods X\% more efficient whilst at the same time the total money stock increased by Y\%, then how do we calculate the changes to the marginal utility of money? When attempting to formulate an answer we will immediately run into the circularity of the definition all over again because any changes to demand for money depend on the change in prices, and the change in prices depends on the change in the demand for money.

Some economists have concluded that the fact that there is a circularity in the relationship between the price level and the marginal utility of money, we must conclude that both are simply undefined on their own and can only be fixed by some exogenous factors, like taxes.  However,  having two variables that are a function of each other in a dynamic system does not necessarily mean that their values are undefined – it depends on the nature of those functions. If their relationships are non-linear then it may be that there is a stable pair of values that corresponds to an attractor, and the system will evolve towards that pair regardless of the initial state.

It is the contention of this paper that the price level and demand for money can be constrained endogenously at least for a model economy. An agent-based model is presented which demonstrates the process.

\section{Proof of convergence of prices}
The proof requires a predefined model of an economy composed of interacting agents with a particular set of rules governing an agent's propensity to spend and an agent's propensity to alter prices when acting as a seller of goods.

\textbf{A simplified economy:} We have an economy made up of agents. All agents are both producers and consumers of a single representative product. Each agent makes a product at a certain number of units per day. They do not consume any of their own produce (a restriction used in many agent-based economic models to ensure trade takes place). Each agent has a shop to sell their own produce. Each agent chooses whatever price they wish to sell their produce and can modify their price at any time. The latest price of their product will always be on display available for all to see. There are no banks and no lending of money or goods takes place. No barter is allowed. There are no taxes. There is no unemployment.

\textbf{Sellers’ pricing policy:} In any economy, one guide that sellers of a good employ to decide on price rises or cuts, is the rate of selling of their good. Certainly, if a good is perishable then the incentive to sell goods before they go off can become extreme and lead to “fire sale” pricing. Even if goods are not perishable, stock storage space can be limited and so full or nearly full storage space will incentivise sellers to lower their prices. Conversely, if their stock rooms are continuously near empty, i.e. their produce is sold almost the instant it is produced, then a rational agent will be tempted to raise their prices.

\textbf{Purchasers’ propensity to spend:} People save for all sorts of reasons, but whatever that reason, their feeling about how satisfied they are with their current quantity of savings will be a function of how much they calculate they could purchase with it. So if the price level fell they would feel like they had more savings and vice versa. We assume that people will have a target level of savings expressed in terms of how much those savings can purchase. So that if someone currently was exactly on target with respect to their savings level but then prices fell then they would become above target and their propensity to spend would increase. Conversely if the price level rose then their perception of their savings would fall below target and their propensity to spend would decrease in order to repair their savings level.

In order to demonstrate that the value of money can be constrained in this model, we must first postulate an equilibrium price level of X and then argue that if the prices were to temporarily fluctuate above X, then mechanisms would ensue that served to lower prices and if the prices were temporarily below X  then mechanisms would ensue that served to raise prices.

So, assuming that such an equilibrium price exists let us first consider the consequences of a temporary fluctuation in that price above that level. The agents will observe the now higher prices – this in turn will cause them to sense that their stock of savings will purchase less than before. This will reduce their propensity to spend. This in turn will lead to rising stock levels in the shops. This in turn will incentivise agents to lower their advertised selling prices.

Now let us consider the consequences of a temporary fluctuation in prices below the equilibrium level. The agents will observe the now lower prices – this in turn will cause them to sense that their stock of savings will purchase more than before. This will increase their propensity to spend. This in turn will lead to falling stock levels in the shops. This in turn will incentivise agents to raise their advertised selling prices.

\section{The agent-based Model}

To demonstrate this idea further, an agent-based computer simulation was written to test the idea.

\section{Initialisation}

A collection of 30 agents was created. Each was endowed with a stock of money as savings. Each is assigned a productivity. I.e. how many units of produce the agent can make on each iteration. Each agent is assigned the same starting price which is a parameter of the simulation.
In order to ensure that agents cannot consume their own produce, it is necessary for each agent to maintain separate records for A) produce made, now available for sale and B) produce purchased, available to be consumed.

\begin{samepage}

At each iteration of the simulation, each agent performs the following actions: 

\begin{itemize}
\item Produce
\item Consume
\item Purchase
\item Modify prices
\end{itemize}

We shall consider each in turn:

\end{samepage}

\textbf{Produce:} Each agent’s quota of “produce made, now available for sale” will increase by the assigned productivity of that agent.

\textbf{Consume:} Each agent’s quota of “produce purchased, available to be consumed” will decrease. This is achieved by multiplying it by a number slightly less than one. This is intended to model the phenomena that richer people tend to consume more than poorer ones and also when your stock of goods to consume diminishes, you are likely to ration your daily consumption ever more strictly.

\textbf{Purchase:} Each agent will interrogate the current price of goods on sale from 3 randomly chosen other agents (that have at least one unit of goods available for sale) and will consider whether or not to make a purchase from the agent offering the lowest price P. In order for a purchase to take place we first of all must meet the criterion that the purchaser must have sufficient money to buy at least one unit of the sellers produce. Secondly a purchase will only go ahead so long as the loss of utility by virtue of losing P units of currency is outweighed by the gain in utility by virtue of adding to our stock of “produce purchased, available to be consumed”. In pseudocode we have:

\begin{algorithm}[H]
  \DontPrintSemicolon

 \For{each agent}{
P  = best\_price\_found\_amongst\_3\_random\_agents()\;
wellbeing\_now = wellbeing(goods\_possessed, savings\_possessed)\;
theoretical\_post\_purchase\_wellbeing = wellbeing(goods\_possessed+1, savings\_possessed-P)\;

  \If{theoretical\_post\_purchase\_wellbeing $>$ wellbeing\_now}
  {
make\_purchase()\;
   }
 }
\end{algorithm}

Each agent’s overall utility is given by the product of two separate utility functions: utility from having goods for consumption and utility from having savings. I.e.

\begin{algorithm}[H]
  \DontPrintSemicolon
  \SetKwFunction{FMain}{wellbeing}
  \SetKwProg{Fn}{Function}{:}{}
  \Fn{\FMain{goods, savings}}{
        ug = utility\_from\_goods(goods)\;
        us = utility\_from\_savings(savings)\;
	
        \KwRet ug $\times$ us\;
  }

\end{algorithm}

The utility function from having goods for consumption is a conventional marginal utility curve with ever decreasing gradient but the utility from having savings is a two step process. Step 1 is to convert the savings into an estimate of how many days' worth of goods could be purchased at the currently prevailing price level and then step 2 is to use a conventional marginal utility curve for that number of days.

\begin{algorithm}[H]
  \DontPrintSemicolon
  \SetKwFunction{FMain}{utility\_from\_savings}
  \SetKwProg{Fn}{Function}{:}{}
  \Fn{\FMain{savings}}{

	cost\_of\_one\_days\_consumption = average\_current\_selling\_price() $\times$ TYPICAL\_GOODS\_MADE\_PER\_DAY\;
        days\_worth\_of\_savings = savings / cost\_of\_one\_days\_consumption\;
	
        \KwRet diminishing\_returns\_utility(days\_worth\_of\_savings)\;
  }

\end{algorithm}

\textbf{Modify prices:} Each agent aims to maintain their prices at the highest possible level subject to two constraints. 1, that the stock levels should not reach (or should very rarely reach) full capacity and 2, The frequency of price changes should be reasonable. This is to emulate the fact that in real life, changing prices are subject to menu costs. In pseudocode:

\begin{algorithm}[H]
  \DontPrintSemicolon

 \For{each agent}{

  \If{days\_since\_last\_price\_change $>$ minimum\_price\_change\_period}
  {
	stock\_growth\_per\_day = goods\_we\_produce\_per\_day - sales\_per\_day\_since\_last\_price\_change\;
	
	\eIf{stock\_growth\_per\_day $>$ 0}
	{
		// storeroom filling up\;
		days\_till\_storage\_full = (MAXIMUM\_STOCK - stock\_for\_sale) / stock\_growth\_per\_day\;
		\uIf{days\_till\_storage\_full $<$ 3}
		{
		selling\_price = selling\_price $\times$ 0.85  // fire sale!\;
		}
		\uElseIf{days\_till\_storage\_full $<$ 15}
		{
		selling\_price = selling\_price $\times$ 0.95\;
		}
		\uElseIf{days\_till\_storage\_full $>$ 90}
		{
			// we can afford to raise price for a short while\;
		selling\_price = selling\_price $\times$ 1.05\;
		}
	}
	{
		// storeroom emptying\;

		\uIf{days\_till\_storage\_empty $<$ 3}
		{
		selling\_price = selling\_price $\times$ 1.1 \;
		}
		\uElseIf{stock\_for\_sale $<$ (MAXIMUM\_STOCK / 2)}
		{
		selling\_price = selling\_price $\times$ 1.05\;
		}
	}
   }

  }

\end{algorithm}

\section{Results}

The average selling price of the agents converges to a fixed level which is independent of the starting price, confirming the existence of a price attractor. 

Fig 2 shows the evolution of the average selling price over time when the starting price was set at a variety of different values.

\begin{figure}[h]
\includegraphics[width=12cm]{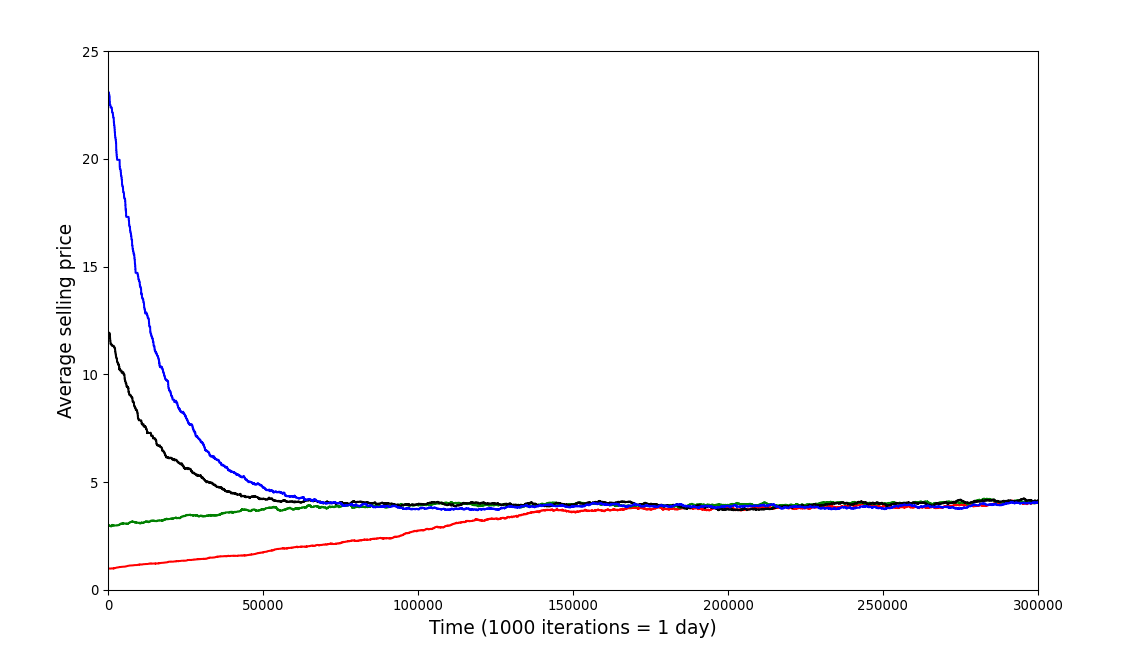}
\caption{Prices converging on the same value independent of starting price}
\end{figure}

\FloatBarrier

\bibliographystyle{apacite}
\bibliography{myrefs}

\end{document}